\documentclass[nofootinbib,twocolumn,longbibliography,aps]{revtex4-1}

\usepackage[dvipsnames]{xcolor} 

\usepackage{color,soul}
\usepackage{blindtext}
\usepackage{float}
\usepackage{epsfig}
\usepackage{dsfont}

\usepackage{amsmath}
\usepackage{braket}
\usepackage{amsfonts}
\usepackage{amssymb}
\usepackage{graphicx}  
\usepackage[dvipsnames]{xcolor}
\usepackage{todonotes}
\usepackage{bbm}
\usepackage{tikz}
\usepackage{lmodern}
\usetikzlibrary{decorations.pathmorphing}
\tikzset{snake it/.style={decorate, decoration=snake}}
\usepackage{fancybox}
\expandafter\ifx\csname package@font\endcsname\relax\else
\expandafter\expandafter
\expandafter\usepackage
\expandafter\expandafter
\expandafter{\csname package@font\endcsname}%
\fi
\DeclareMathOperator{\sech}{sech}

\RequirePackage[colorlinks,citecolor=blue,urlcolor=magenta,linkcolor=blue]{hyperref}

\newcommand{\be}{\begin{equation}}
\newcommand{\ee}{\end{equation}}
\newcommand{\bea}{\begin{eqnarray}}
\newcommand{\eea}{\end{eqnarray}}

\def\ket#1{|#1\rangle}                    

\begin{document} 
\title{A nested sequence of  inequivalent Rindler vacua :  Universal  Relic Thermality of Planckian origin}
\author{Kinjalk Lochan}
\email{kinjalk@iisermohali.ac.in}
\affiliation{Department of Physical Sciences, Indian Institute of Science Education \& Research (IISER) Mohali, Sector 81 SAS Nagar, Manauli PO 140306 Punjab India.}
\author{T. Padmanabhan}
\email{posthumously}
\affiliation{IUCAA, Post Bag 4,  Pune University Campus, Ganeshkhind 400011 India.}

\begin{abstract}
The  Bogoliubov transformation connecting the standard inertial frame mode functions  to  
 the standard  mode functions defined in the Rindler frame $R_0$, leads to the result that the inertial vacuum appears as a thermal state with temperature $T_0=a_0/2\pi$ where $a_0$ is the acceleration parameter of $R_0$. We construct an infinite  family of nested Rindler-like coordinate systems $R_1, R_2, ...$ within the right Rindler wedge, with time coordinates $\tau_1, \tau_2, ...,$ and acceleration parameters $a_1, a_2, ...$ by shifting the origin along the inertial $x$-axis by amounts $\ell_1, \ell_2,\cdots$. We show that, apart from the inertial vacuum, the \textit{Rindler vacuum} of the frame $R_n$ also appears to be  a thermal state in the frame $R_{n+1}$ with the temperature $a_{n+1}/2\pi$. \textit{In fact, the Rindler frame $R_{n+1}$  attributes to all the Rindler vacuum states of $R_1, R_2, ... R_n$, as well as to the inertial vacuum state, the same temperature $a_{n+1}/2\pi$.}   We further show that our result is discontinuous in an essential way in the coordinate shift parameters. For a Rindler frame $R_i$, this thermality {\it turns on} with smallest non-zero $\ell_i$ allowed in the  semiclassical framework and remains insensitive to $(\ell_i,a_{i-1})$ thereafter, indicating its universal Planckian origin. Similar structures can be introduced in the right wedge of any spacetime with bifurcate 
Killing horizon, like, for e.g., Schwarzschild spacetime.  Apart from providing unsuppressed observables capturing Planck scale effects,  these results  have important implications for quantum gravity when flat spacetime is treated as the ground state of quantum gravity.
Furthermore, a frame with the shift $\ell$ and the corresponding acceleration parameter $a(\ell)$ can be thought of as a Rindler frame which is  instantaneously comoving with the Einstein's elevator moving with a variable acceleration. Our result suggests that the quantum temperature associated with such an Einstein's elevator is the same as that defined in the comoving Rindler frame.  The implications of  these results are wide ranging, from having a definitive signature of Planck shifts in the horizon to the existence of a new set of observers in black hole exterior having thermodynamic description of the horizon they perceive.
 
\end{abstract}

\maketitle

\section{Introduction}
\begin{figure}[h]
    \begin{center}
        \includegraphics{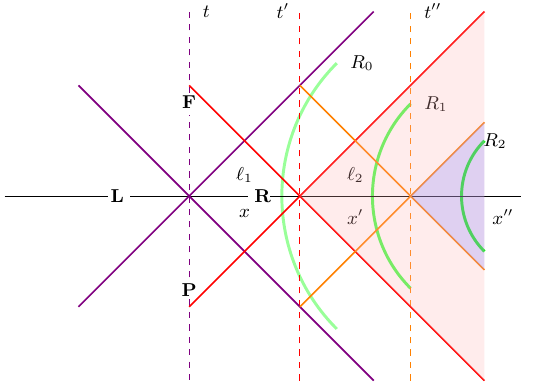}

    \end{center}
    \caption{Translated Rindler trajectories inside other Rindler wedges.} 
    \label{fig:RindlerRindler01}
\end{figure}

It is a common understanding that the inertial observers related by translations are equivalent in the sense that there is no non-trivial Bogoliubov transformation between their vacua. If they were, simple translations would have generated particles. However, it is equally interesting fact that each inertial frame also contains in its inside a Rindler description, in which the inertial vacuum appears as a thermofield double state \cite{Fabbri:2005} over the two Rindler wedges,
\bea
|\textrm{0}\rangle_M = N \prod _{\omega} \sum _{n} e^{-\frac{n \beta  \hbar\omega}{2}}~|n_{\omega}^{\rm L}\rangle \otimes |n_{\omega}^{\rm R}\rangle.
\eea
If the inertial vacua under translations are all equivalent, are their respective Rindler descriptions  also equivalent under inertial spatial translations?   We analyze the question of equivalence of different non-inertial frames constructed on different points in Minkowski space, related by translations. One can visualize from this construction that many of the Rindler wedges will have non trivial overlap with the Rindler wedges of some other points and for spatial translations this will generate a sequence of Rindler frames  nested under one another, see Fig.(\ref{fig:RindlerRindler01}). If we set up different acceleration trajectories in such two different frames, they will explore different causal regions. Therefore, it would be worth pondering over the question whether the experiences of such two Rindler frames are also completely equivalent  w.r.t. one another, because of their seed inertial frames being completely equivalent to each other.  We investigate this question through the study of quantum fields in two Rindler frames and ask if there exists a unique vacuum state across all Rindler frames irrespective of their location or whether spatial translation generates inequivalent Rindler vacua.

This setting has an equivalence in terms of an  observer moving under a space dependent weak gravitational field expressed in term of the line element  \cite{Rindler:1966zz, Padmanabhan:2009gm}
\bea
ds^2 = -(1+  2\phi (x)) dt^2 + dx^2 \approx -(1+  \phi (x))^2 dt^2 + dx^2.
\eea
The equivalence principle suggest that at each point $x$ the local gravity is equivalent to an accelerating frame determined by the local potential $\phi(x)$. Thus as the particle moves in this potential along the spatial direction its equivalence frame will be a sequence of nested Rindler frames as described above, to the leading order in the departure from the base point $x_0$, 
\bea
ds^2 &= &-[1+  \phi (x_0) + \phi'(x_0) (x-x_0)]^2dt^2 + dx^2 \nonumber\\
&=& -(1  + \alpha \tilde{x})^2d\tilde{t}^2 + d\tilde{x}^2,
\eea
where $\tilde{t} = t/(1+\phi(x_0))^{1/2}, \tilde{x}=x-x_0$ and $\alpha =\phi'(x_0)/(1+\phi(x_0)).$

We can also consider the generalization of this  concept for the {\it Einstein's elevator} moving along the $x$-axis with a variable acceleration $a(\ell)$  for their metric 
when the elevator has moved a distance $\ell$. We can now introduce a sequence of Rindler frames $R_1, R_2, ...$ each with acceleration parameter $a_1, a_2, ...$ such that the Rindler frame $R_j$ is instantaneously comoving with the  elevator, (i.e, moving with the same velocity \textit{and acceleration} $a_j = a(\ell_j)$) when it is located at $x= \ell_j$. 
%
This construction also has an interesting parallel with the use of a sequence of comoving \textit{inertial} observers in special relativity. Recall that, to  study the time dilation of an accelerated clock $C$ moving with the velocity $v(t)$ with respect to a global, inertial, lab frame  $S$ (the coordinate $t$ represents the inertial time coordinate of the lab frame), we introduce  a sequence of \textit{instantaneously comoving} inertial frames $S_1, S_2, ...$ connected to $S$ by Lorentz boost with velocities $v_1, v_2, ...$. The comoving inertial frame $S_j$ will have the same speed, $v_j \equiv v(t_j),$ as the clock at some instant $t=t_j$ and its origin will coincide with the location of the clock at $t=t_j$. An ideal clock is \textit{postulated} to be the one for which the proper time lapse $\Delta \tau$ will be the same as the time lapse in the  instantaneously comoving inertial clock. Therefore we can use the relation $d\tau = [1 - v^2(t)]^{1/2} dt$ for ideal clocks in \textit{accelerated} motion (see chapter 16 in \cite{Misner:1973prb}). By our postulate, the flow of time of ideal clocks is unaffected by e.g, the  acceleration and depends only on the instantaneous velocity of the comoving inertial frame. Just as a sequence of comoving inertial frames are introduced to study the clock motion with \textit{variable speed} in special relativity, in the nested non inertial frame setting, we have now introduced a sequence of comoving Rindler frames to study an elevator moving with \textit{variable acceleration} along the $x$-axis.

Thus, the consideration of the quantum fields by observers moving in weak gravitational fields will also ascribe if the equivalence principle is respected at the quantum level through the instantaneous acceleration, and it remains insensitive to the other details of the gravitational potential.

In order to address these questions systematically, let us consider a massless scalar field residing in the 1+1 dimensional flat spacetime described using the standard inertial coordinate chart $(t,x)$. The null lines $t=\pm x$, divide the spacetime into four wedges $R, F, L, P$ (Fig.\ref{fig:RindlerRindler01}). We introduce the Rindler coordinate frame $R_0(a_0)$, which depends on the acceleration parameter $a_0$, through the coordinates transformation  
$
a_0x=e^{a_0\xi_0}\cosh a_0\tau_0; \qquad  a_0t=e^{a_0\xi_0}\sinh a_0\tau_0. 
$
Since $\tau_o$ parametrizes the integral curves of the boost generator, the spacetime is stationary in the Rindler coordinate time $\tau_0$. One can now expand the scalar field in terms of modes which are positive/negative  frequency with respect to either the inertial time coordinate (like $\exp(\pm i \Omega t)$) or positive/negative frequency with respect to the Rindler time co-ordinate (like $\exp(\pm i\omega_0 \tau_0)$). These two sets of modes are related by a Bogoliubov transformation  leading  to the well known result that the inertial (Minkowski)  vacuum state $\ket{0}_M$ will appear to be thermally populated by Rindler particles at a temperature $T_0 = a_0/2\pi$ \cite{Unruh1976, Birrell1982, Mukhanov-Winitzky}. More precisely, the expectation value of the Rindler number operator will be thermal in the inertial vacuum state.

Consider now another Rindler-like wedge which is completely contained within the standard wedge $R_0$. This new wedge can be obtained by shifting the origin of coordinates to the right by a distance $\ell_1$ so that the null horizons now emanate from the event ($t=0,x=\ell_1$),  as shown in Fig. (\ref{fig:RindlerRindler01}). One can again introduce a Rindler-like coordinate system ($\tau_1,\xi_1$) in the new Rindler-like wedge $R_1$ with the acceleration parameter $a_1$  by the coordinate transformations: 
$
a_1 (x-\ell_1) =  e^{a_1\xi_1}\cosh a_1\tau_1;  \qquad  a_1  t =  e^{a_1\xi_1}\sinh a_1\tau_1.
$ 
The inertial vacuum, $\ket{0}_M$, is of course translation invariant; therefore it is obvious that $\ket{0}_M$ will appear to be thermally populated with temperature $T_1 = a_1/2\pi$ when viewed within the wedge $R_1$. But the interesting question to ask is: {\it How does the Rindler vacuum $\ket{R_0}$ of the right wedge appear in the region $R_1$? }

By eliminating the inertial coordinates $(T,X)$ appearing in $R_0$ and $R_1$ co-ordinates, we can express the coordinates $(\tau_1,\xi_1) $ directly in terms of $(\tau_0,\xi_0)$. It then straightforward to show that  the positive/negative frequency modes $\exp(\pm i\omega_0 \tau_0)$ are related by a \textit{non-trivial} Bogoliubov transformation to the positive/negative modes $\exp(\pm i \omega_1 \tau_1)$. Therefore, the notion of particles in $R_0$ and $R_1$ are non-trivially different. 

As we shall  see, quite remarkably, the Rindler vacuum  in $R_0$ appears to be \textit{thermally populated in $R_1$ with the temperature $T_1 = a_1/2\pi$!} This result has the following, very interesting, features: 
\begin{enumerate}
  \item The temperature depends only on the (``local'') acceleration parameter of $R_1$ and has no memory of the fact that $R_0$ itself is defined with an acceleration parameter $a_0$ with respect to the inertial coordinates.

\item The Minkowski vacuum $\ket{0}_M$ will \textit{also} appear to be thermally populated with the same temperature $a_1/2\pi$ in the sub-wedge $R_1$. In other words, \textit{both} the Minkowski vacuum as well as the Rindler vacuum of the right wedge appear to be thermally populated with the same temperature in the sub-wedge $R_1$. 
 
 \item The result turns out to be  independent of the shift $\ell_1$ in a rather subtle way. It turns out that if $\ell_1=0$, then the effect disappears and the Rindler vacuum of $R_0$ will not contain any particles with respect to $R_1$. But, if $\ell_1 \ne 0$, the effect kicks in (with thermal population at temperature $a_1/2\pi$) \textit{however small the shift $\ell_1$ is}.  {\it Since no event can be specified to an accuracy better than Planck length, this resultant discontinuous emergence of thermality, even at $\ell \sim \ell_{QG}$ is intriguing.} Comparison of number spectrum and correlators thereof, in such Rindler frames, provides a direct estimator of Planck scale effects which is not heavily suppressed in the semiclassical regime, but is rather as strong as the Unruh effect.
 \end{enumerate}

  Obviously this exercise can be continued indefinitely by introducing a nested series of Rindler-like frames $R_1 (\ell_1, a_1), R_2 (\ell_2, a_2) \cdots R_n (\ell_n, a_n)\cdots$ with two parameters (shift $\ell_n$ and acceleration $a_n$) characterizing the $n$-th nested Rindler wedge. On repeating the calculation, we find that all the Rindler vacua corresponding to the frames $R_0,R_1,\cdots R_n$ will appear to be thermally populated in $R_{n+1}$ with the temperature $a_{n+1}/2\pi$. In fact, one can consider a continuum of nested Rindler frames $R(\ell)$, parametrized by the shift $\ell$ and a function $a(\ell)$ which gives the acceleration parameter $a(\ell)$ of the frame $R(\ell)$ with $0<\ell<\infty$.
  
To avoid possible misunderstanding, we stress the following fact. \textit{In order to study the Bogoliubov transformation and study of relation among the Fock bases between two valid co-ordinate frames in a region of spacetime, we do not have to introduce the notion of observers or particle detectors \textit{per se}.} It is well known that what the particle detectors see in a quantum state can be quite different from the conclusions we will obtain via Bogoliubov transformations \cite{Davies1996, Sriramkumar:1999nw}. Throughout this paper, we will be interested \textit{only} in the formalism of QFT through mode functions and Bogoliubov transformations and will not discuss detector response. The detector response in this set up is discussed in \cite{Sahota:2024thk}, essentially giving the results  obtained through the Bogoliubov computations.
\\



\section{Outline of the calculation}
The mode functions of the massless scalar field in $R_0$ and $R_1$ are simply plane waves in their respective co-ordinates and  the Bogoliubov coefficients can  directly be computed from their overlap (see Supplementary Material). However, in order to illustrate this with clarity, we adopt another formal way as follows. 
The mode functions between the inertial ($M$) and the standard Rindler right wedge ($R_0$) are related through the Bogoliubov coefficients  $\alpha_{\omega \omega'} (a_0), \beta_{\omega \omega'} (a_0)$ as
$
u_{\omega}^{R_0} = \sum_{\omega'} [ \alpha_{\omega, \omega'} (a_0) u_{\omega'}^{M} + \beta_{\omega, \omega'} (a_0)  (u_{\omega'}^{M} )^*].
$
Let $M'$ be another inertial frame $(t',x')$ related with $M (t,x)$ by a 
 spatial translation $ x'=x-\ell_{1}$. 
Let $R_1$ be the Rindler frame in the right wedge of $M'$ defined using an acceleration parameter $a_1$ and let $\alpha_{\omega, \omega'} (a_1), \beta_{\omega, \omega'} (a_1)$ be the Bogoliubov coefficients between $M' \leftrightarrow R_1$.
Using  $u_{\omega}^{M'} = e^{-i \omega \ell_{1}}u_{\omega}^{M}$ , the mode functions for $R_1$  can be written as
$
u_{\omega}^{R_1} = \sum_{\omega'} [\alpha_{\omega, \omega'} (a_1)e^{-i \omega \ell_{1}} u_{\omega'}^{M} + \beta_{\omega, \omega'} (a_1)e^{i\omega \ell_{1}} (u_{\omega'}^{M} )^*].$
The Rindler wedge  $R_1$ is completely contained within  $R_0$; see Fig. (\ref{fig:RindlerRindler01}). So   
  standard relation between
the modes $u_{\omega}^{M},$ --- defined on the whole Cauchy surface (and hence inside $R_1$ as well) with specific coordinate representation in each of the four wedges \cite{Gerlach:1988}---   and  $u_{\omega}^{R_0} $; 
$
u_{\omega'}^{M} = \sum_{\omega''} [ \alpha^*_{\omega'', \omega'} (a_0)  u_{\omega''}^{R_0} - \beta_{\omega'', \omega'} (a_0)  (u_{\omega''}^{R_0} )^*],
$
holds inside $R_1$ without any reference to the outside region. This allows us to relate the modes of $R_1$ and $R_0$ directly. After some straightforward algebra we find that
the Bogoliubov coefficient between the the two Rindler frames $R_0$ and $R_1$ is given by
\begin{widetext}
\bea
\tilde{\beta}_{\omega, \omega''}=
\sum_{\omega'}\left(-\alpha_{\omega, \omega'} (a_1)e^{-i \omega' \ell_{1}} \beta_{\omega'', \omega'} (a_0) + \beta_{\omega, \omega'} (a_1)e^{i \omega'  \ell_{1}} \alpha_{\omega'', \omega'} (a_0)  \right). \label{EffectBeta1}
\eea

When the two frames have the same acceleration parameter and the shift is zero, this expression gives $\tilde{\beta}_{\omega, \omega''}=0$ following the identity the Bogoliubov coefficients follow \cite{Mukhanov-Winitzky},  which remains true even when the 
 two accelerations are different but the shift is zero.  To see this, we need to use the explicit form of the  Bogoliubov coefficients between the $j-$th inertial and Rindler pair (with $j=0,1$):
\bea
\alpha_{\omega, \omega'}/\beta_{\omega, \omega'} (a_j)=\pm\frac{1}{2 \pi a_j}e^{\pm\frac{\pi \omega}{2a_j}}\sqrt{\frac{\omega}{\omega'}}e^{-i\frac{\omega}{a_j}\log{\frac{\omega'}{a_j}}}\Gamma\left( \frac{i \omega}{a_j} \right),
\eea
leading to
\bea
\tilde{\beta}_{\omega, \omega''}=\frac{1}{4 \pi^2} \frac{\omega}{\sqrt{a_0 a_1}} \left( \frac{ \left| \Gamma\left( \frac{i \omega}{a_1} \right)\right|^2}{a_1} \delta \left( \frac{\omega}{a_1}+ \frac{\omega''}{a_0}  \right)- \frac{ \left| \Gamma\left( \frac{i \omega''}{a_0} \right)\right|^2}{a_0}\delta \left( \frac{\omega}{a_1}+ \frac{\omega''}{a_0}  \right)\right) e^{i  \frac{\omega}{a_0}\log{\frac{a_0}{a_1}}} =0, \label{ZeroAcceleration2}
\eea
as the two delta functions individually vanish for their positive arguments. This is to be expected because, in this case, the two time coordinates are related by a rescaling and positive frequency modes map to positive frequency modes (see Appendix A in the Supplementary Material).
On the other hand, if the two accelerations are different then for non-zero positive shift, $\ell_{1} > 0$ we get the nontrivial result:
\bea
\tilde{\beta}_{\omega, \omega''}=-\frac{1}{ 2\pi^2} \frac{\sqrt{\omega\omega''}}{a_0 a_1}  \left(\frac{a_0}{a_1}\right)^{i \frac{\omega''}{a_0}}(a_1 \ell_{1})^{i\left(\frac{i \omega}{a_1}+\frac{i \omega''}{a_0} \right) }\Gamma\left[ \frac{i \omega}{a_1} \right]\Gamma\left[ \frac{i \omega''}{a_0} \right]\Gamma\left[-\left( \frac{i \omega}{a_1}+\frac{i \omega''}{a_0} \right) \right] \sinh{\frac{\pi \omega''}{a_0}}. \label{EffBogBeta1}
\eea
\end{widetext}
Therefore, the vacuum state of $R_0$ frame will appear to be populated by particles of the $R_1$ frame with the number of 
 particles being 
  \bea
 N_{\omega}&\equiv& \langle 0_{R_{0}}|\hat{N}_{\omega}| 0_{R_{0}}\rangle = \int d \omega'' |\tilde{\beta}_{\omega, \omega''} |^2  \nonumber\\
 &=&\frac{1}{4 \pi a_1} \frac{1}{\sinh{\frac{\pi \omega}{a_1}}}\int_0^{\infty} \frac{d \omega''}{a_0} \frac{\sinh{\frac{\pi \omega''}{a_0}}}{\left( \frac{\omega}{a_1}+\frac{\omega''}{a_0} \right)\sinh{\left( \frac{\omega}{a_1}+\frac{\omega''}{a_0} \right)}}.\nonumber\\
 \label{ParticleRindlerFull}
\eea
Simplifying this expression, we find that :
\bea
N_{\omega}\approx \frac{1}{2 \pi a_1}\frac{\pi \delta(0)}{e^{\frac{2 \pi \omega}{a_1}}-1},
\label{ParticleRindler}
\eea
upto some subdominant finite correction terms (see  Appendix B in the Supplementary Material for details). So the number density of particles has a dominant thermal form similar to the Unruh effect, with the departures from thermality (infinitely) suppressed by the volume measure. In fact, all correlations of the spectral density follow a thermal profile 
\bea
 \langle 0_{R_{0}}| \hat{N}_{\omega_1}\hat{N}_{\omega_2}| 0_{R_{0}}\rangle \approx \frac{1}{(2 \pi a_1)^2}\frac{\pi \delta(0)}{e^{\frac{2 \pi \omega_!}{a_1}}-1}  \frac{\pi \delta(0)}{e^{\frac{2 \pi \omega_2}{a_1}}-1}, \label{CorrelatorRindler}
 \eea
as a consequence (see Appendix B in the Supplementary Material for detailed discussions).
A few points are noteworthy here. 
\begin{itemize}
\item  Since $\ell_1$ appears only through an overall phase in Eq. (\ref{EffBogBeta1}), the thermal density Eq. (\ref{ParticleRindler}) is evidently  independent of $\ell_{1}$ (with an understanding that $\ell_1 \neq 0$) even if we include sub-dominant terms upto all orders, see Eq. (\ref{ParticleRindlerFull}). So is the case with the correlators as well, Eq.(\ref{CorrelatorRindler}). The parameter $\ell_j$ effectively signifies the amount of non-overlap between two consecutive Rindler wedges $R_j$ and $R_{j+1}$.  As we can see from  Fig.(\ref{fig:RindlerRindler01}),  two non-consecutive Rindler frames $R_k$ and $R_j$ with $k<j$ are also related by an effective shift $\ell_{kj}\equiv \ell_{k+1}+...+\ell_{j-1}+\ell_{j}$, the vacuum of $R_k$ also appears thermal to $R_j$ with no dependence on $\ell_{kj}$. Further, the number density spectrum or correlators  do not contain any reference to the acceleration $a_0$ of $R_0$ frame (or $a_k$ of the frame $R_k$ for a non-consecutive case) even with the subleading corrections included (See Appendix B in the Supplementary Material). This means that  the vacua of all the preceding Rindler frames in the nested Rindler structure will appear the same, as far as correlations are concerned,  despite being inequivalent ($\tilde{\beta}_{\omega, \omega''}$ between any two frames is non-zero, Eq. (\ref{EffBogBeta1})).  {\it As a consequence, we have a family of  inequivalent vacua corresponding to $R_k$s,  all of which appear exactly the same, even after including all sub-dominant corrections at all orders, i.e., thermal with a temperature $T_j \propto a_j$ in the frame $R_j$.} 

\item  Most importantly, as long as $\ell_j$ between two Rindler frames is non-zero,  but howsoever small, the emergence of the effective thermality will {\it turn on}. Since, in the semiclassical analysis,  $\ell_j$  can go down all the way to the smallest possible value $\ell_{QG}$, possibly the Planck length (where quantum gravity effects are expected to modify the semiclassical picture considerably), {\it the comparison of particle spectrum or even correlators in the vacua of two adjacent but Planck length apart (in terms of the null rays they asymptote to) Rindler frames ${\cal F}(\omega_1,\omega_2) \equiv \langle 0_{R_{i+1}}| \hat{N}_{\omega_1}\hat{N}_{\omega_2}| 0_{R_{i+1}}\rangle -\langle 0_{R_{i}}| \hat{N}_{\omega_1}\hat{N}_{\omega_2}| 0_{R_{i}}\rangle$
provides a remarkably direct and robust marker to spot even a Planck scale mismatch in the overlap of their causal domains.} Typically such Planck scale effects are heavily suppressed at low energies in any (semi-) classical analysis, but  in the nested Rindler frame structure they appear quite robustly  (i.e., as strong as the Unruh effect). The fact that after turning on with a Planck scale shift,  the residual thermality remains exactly the same (including all higher order corrections) for the all the {\it inequivalent vacua} of all previous nested Rindler frames and is independent of their individual characteristics $(a_{i-1},\ell_i)$s  afterwards, clearly illustrates that it is indeed a relic of the Planck scale effect which provides a base thermal quanta to all previously nested Rindler frames.

\item Lastly, the density of the particle excitations in the Rindler vacuum is exactly half the number density of the of excitations had the vacuum been inertial. Nevertheless, the Boltzmann factor (ratio of excitations across different frequencies)  remains thermal with the temperature $T_j \propto a_j$. This conclusion can also be arrived at from  the periodicity of the correlation functions in the vacuum of $R_0$ as perceived from the frame $R_1$ (see Appendix C in the Supplementary Material).

\end{itemize}
This property of relic thermality can be expected to persist for any frame which remains in a region of spacetime which is not maximally extended. 
For instance, in a wide variety of 1+1 dimensional curved spacetimes with a bifurcate Killing horizon one can introduce global Kruskal-like coordinates with the past and future horizons, thereby separating the spacetime into $R,F,L,P$ regions and a nested Rindler structure therein.  We will now illustrate the feature of reminiscent Planck scale thermality  in the case of 1+1 dimensional Schwarzschild spacetime. 
\section{Generalization: Example of 1+1 dimensional Schwarzschild Exterior}

To exploit the conformal invariance and define the mode functions, it is useful to work in 
the ``tortoise null co-ordinates'', outside the horizon at $r=r_g$. In these coordinates,  the Schwarzschild metric is:
\bea
ds^2 =-\frac{r_g}{r(\tilde{u},\tilde{v})}\exp\left( 1 - \frac{r(\tilde{u},\tilde{v})}{r_g}\right)e^{-\frac{(\tilde{u}- \tilde{v})}{2 r_g}}d\tilde{u} d \tilde{v}.
\eea
The relation between the global, Kruskal null co-ordinates $(u,v)$, spanning the maximal extension of the Schwarzschild spacetime and the tortoise null co-ordinates is given as
\bea
u &=& -2r_ge^{-\frac{\tilde{u}}{2 r_g}},\nonumber\\
v &=& 2r_g e^{\frac{\tilde{v}}{2 r_g}}.
\eea
The line element in  the Kruskal null co-ordinates has the form:
\bea
ds^2 =-\frac{r_g}{r(u,v)}\exp\left( 1 - \frac{r(u,v)}{r_g}\right)d u d v.
\eea
These co-ordinates are valid everywhere in the causal diagram; see Fig. (\ref{fig:BlackHoleRindler}). For a massless scalar field, the functions $ \phi_{\omega} (u) = e^{-i \omega  u}  ,  e^{-i \omega v}$
define an orthonormal set of modes  everywhere is the causal diagram.  Similarly,  in the tortoise co-ordinates which are defined in the exterior of the black hole, we have another set of orthonormal modes, $\phi_{\omega} (\tilde{u}) = e^{-i \omega  \tilde{u}}  ,  e^{-i \omega \tilde{v}}$.
The ``outgoing'' mode $e^{i \omega u}= e^{-i \omega (T-R)}$,  defines a positive frequency mode with respect to co-ordinate $T$ which is the proper time in the near horizon region because the metric near the horizon, in the  Kruskal coordinates becomes
\bea
ds^2 \rightarrow -dudv = -dT^2 + dR^2,
\eea
for $u = T-R, v =T+R$.
Similarly, the mode $e^{-i \omega  \tilde{u}}= e^{-i \omega(t-r_*)}$ is a positive frequency out-going mode function with respect to co-ordinate $t$ which is the proper time of in the asymptotically flat region of the spacetime.
Writing 
$\tilde{u} = t-r_*, \tilde{v} = t+r_*$
we obtain
\bea
T &=& 2r_g e^{\frac{r_*}{2r_g}}\sinh{\frac{t}{2r_g}},\\
R &=& 2r_g e^{\frac{r_*}{2r_g}}\cosh{\frac{t}{2r_g}}.
\eea
 Since the coordinate transformation between the Kruskal and the stationary (tortoise) frames are exactly similar to that between the Minkowski and Rindler null-cordinates (see Appendix A in the Supplementary Material), the {\it Kruskal vacuum} will exhibit a  thermal population of particles defined using the mode functions of the tortoise null coordinate system with the temperature $T=1/4\pi r_g$ \cite{Hawking1974,Fabbri:2005, gravitation}. The parameter $r_g$ plays the role of the acceleration parameter of the Rindler frame in flat spacetime.

\begin{figure}[h]
    \begin{center}
        \begin{tikzpicture}[scale=1.25]   
            \draw[black] (1.5,-1.5) -- (3,0);
            
            \draw[black, thin] (3,0) -- (1.5,1.5);
         
            \draw[black] (-1.5,-1.5) -- (-3,0) -- (-1.5,1.5);
             \draw[orange,snake it] (-1.5,1.5) -- (1.5,1.5);        
               \draw[orange,snake it] (-1.5,-1.5) -- (1.5,-1.5);

            \draw[magenta,thick,dashed, bend left=45] (1.5,-1.5) to (1.5,1.5);
            \draw[cyan,thick,dashed, bend left=45] (2,-1) to (2,1);
        
            \draw[violet, thick] (-1.5,-1.5) -- (1.5,1.5);
            \draw[violet, thick] (-1.5,1.5) -- (1.5,-1.5);

             \draw[blue, thick] (-0.5,-1.5) -- (2,1);
             \draw[blue, thick] (-0.5,1.5) -- (2,-1);

            \node[label=right:$\mathcal{J}_R^-$] (5) at (2,-1) {};
             \node[label=right:$\mathcal{J}_R^+$] (9) at (2,1) {};
             \node[label=above: $i^{+}$] at (1.5,1.5) {};
             \node[label=below: $i^{-}$] at (1.5,-1.5) {};

               \node[label=above: $${u=0}$$] at (-1.4,-1.3) {};
               \node[label=above: $${u=$-\Delta$}$$] at (-0.5,-1.3) {};

              \node[label=below: $${v=0}$$] at (-1.4,1.3) {};
               \node[label=below: $${v= $\Delta$}$$] at (-0.5,1.3) {};
  
   \fill[cyan!70,nearly transparent]  (2,-1) -- (1,0) -- (2,1) --(3,0) -- cycle;

        \end{tikzpicture}
    \end{center}
 \caption{The observers who use mode functions as  $e^{-i \omega \tilde{u}'}$ spend their entire trajectory in the shaded region and end up on future and past null infinities.}
    \label{fig:BlackHoleRindler}
\end{figure}

We now introduce a shift in the origin and a new parameter $r_g'$ through the coordinate transformations:
\bea
 2r_g e^{\frac{r_*}{2r_g}}\sinh{\frac{t}{2r_g}}&=& 2r'_g e^{\frac{r'_*}{2r'_g}}\sinh{\frac{t'}{2r'_g}},\\
 2r_g e^{\frac{r_*}{2r_g}}\cosh{\frac{t}{2r_g}}-\ell &=& 2r'_g e^{\frac{r'_*}{2r'_g}}\cosh{\frac{t'}{2r'_g}}.
\eea
This transformation is the exact analog of the co-ordinate transformation connecting $R_0\rightarrow R_1$ in the flat spacetime case with identification $(2a_0,2a_1\rightarrow r_g^{-1},r_g'^{-1})$ (see Appendix A in the Supplementary Material), which is equivalent to a transformation between the Kruskal null and these new null co-ordinates $(\tilde{u}', \tilde{v}')$ as 
\bea
u &=& -2r'_ge^{-\frac{\tilde{u'}}{2 r'_g}} -\ell \nonumber\\
v &=& 2r_g e^{\frac{\tilde{v'}}{2 r'_g}}+\ell.
\eea
Therefore,  the full range of the new null co-ordinates $ -\infty <\tilde{u}', \tilde{v}'<\infty$  only covers the region $u\in[-\infty, -\ell], v \in [\ell, \infty]$, which is a region contained in the exterior region (the right wedge of the Kruskal frame). This is a subspace of the manifold covered by the tortoise coordinates. 
The line element in terms of the new co-ordinates is given as
\bea
ds^2 = - \frac{r_g e^{\left(-\frac{r(r',t')}{r_g} + \frac{r'}{r'_g}  \right)}}{r(r',t')}\left( \frac{r'}{r_g'} -1 \right)\left[dt'^2 -\frac{dr'^2}{\left( 1-\frac{r_g'}{r'} \right)}\right], \label{NewConformal} \nonumber\\
\eea

Clearly, functions $e^{-i \omega \tilde{u}'}$ are valid positive energy mode functions with respect to co-ordinate $t'$, in the full range of the new co-ordinates $\tilde{u'},\tilde{v'}$  but are valid only in a causal patch contained within the exterior region. Therefore,  the asymptotic past and asymptotic future of these co-ordinates end up on respective null boundaries. Such Rindler like hyperbolic trajectories in curved spacetime have been studied previously in various contexts \cite{Rindler:1960, Gaureau:1969, Paithankar:2019, Kerachian:2020}, but  here we provide their quantum field theoretic treatment. Since the co-ordinate transformation between these two frames is exactly similar to those between the nested Rindler, we will essentially get the similar Bogoliubov coefficients and ultimately the similar occupation number expression

\bea
N_{\omega}= \frac{ 1}{ 2\pi }\left[ \frac{\pi \delta(0)}{e^{4 \pi \omega r_g'}-1} \right],\label{ParticleBoulware}
\eea
where, again, the subleading terms vanish in front of $\delta(0)$. Hence, analogous to the Rindler case, the Boulware vacuum will appear thermal in the frame $(\tilde{u}', \tilde{v}')$
 at the temperature decided by $r_g'$ without any reference to $r_g$. Further, just like the Rindler case, in this new frame with $\ell \neq 0$,  the number density in the  Boulware vacuum turns out as half of that in the Kruskal vacuum, while still 
maintaining the thermal distribution across the modes.

 Moreover, if the location of the horizon and hence the area of the black hole is perceived to be even marginally different  in two frames, say due to quantum gravity effects e.g. \cite{SussKind:1993}, they will get thermally connected\footnote{(Incidentally, a similar  feature  also arises for the inertial observers in the case of dilatonic black holes in 2-dimensions \cite{Lochan:2016nbs, Lochan:2016cxt}}. Given the role of Rindler horizon as a local thermodynamic surface, it appears as the accelerated observers in the exterior regions could have their own description of horizons. However in the Schwarzschild exterior, these observers are not necessarily the uniformly accelerated observers in the exterior \cite{Paithankar:2019}. Hence there is a possibility of having  a set of non uniform non static observers in curved spacetime which also may find horizons thermodynamic.
\section{Other Implications}
\begin{itemize}
\item Since the family of vacua discussed above are  inequivalent, yet having  similar particle content (and few lower correlators), the results also have intriguing implications for quantum gravity when we think of flat spacetime as the ground state of quantum gravity (QG) with all matter fields residing in their respective vacuum states. To make this picture consistent, it is necessary to choose the appropriate vacuum state for the matter state and the natural state seems to be the global inertial vacuum $\ket{0}_M$. The question arises as to the role of the Rindler vacua (the usual one plus  the infinite copies of the nested Rindler vacua  discussed in this work) in the ground state description of QG. 

\item In this connection, it must be recalled that the standard Rindler vacuum and the inertial vacuum live in different Hilbert spaces which are unitarily inequivalent \cite{Gerlach:1989rz}.  In a similar spirit, infinite family of the Rindler vacua discussed in this work are unitarily inequivalent and should rather be identified through their parameters of shift $\ell_j$. The situation is somewhat analogous to the $\theta$-vacua in Yang-Mills theories though it is difficult to come up with the instanton solutions connecting these vacua etc. \cite{Belavin:1975fg, Callan:1976je}. The rich structure of flat spacetime is very likely  to 
have implications  in the full theory of QG, especially since ground states play a crucial role in any QFT. Further, the inequivalent vacua description may also be of relevance for cosmological studies, particularly for the de Sitter spacetime where causal patches of each point are different and not fully overlapping, despite the underlying space having a translational symmetry due to its maximal symmetric character, much in the spirit of the inertial and the nested Rindler structure, discussed in this letter. These issues are presently under investigation and will be reported separately.

\end{itemize}

\section{Conclusion}
We have shown that any Rindler frame in the nested sequence of Rindler frames finds vacua of all its causal superspaces thermally populated, see Table~(\ref{Table1}). Our analysis demonstrates that the instantaneously comoving Rindler frame $R_j$ with acceleration $a_j$ will attribute a temperature $a_j/2\pi$ to the vacuum state of all the Rindler frames $R_k$ with $k<j$. In this sense, the comoving Rindler frames only care about the instantaneous acceleration and not, for e.g., the variation of the acceleration. This is reminiscent of the fact that ideal clocks only care about the speed and not about the change in the speed. 
The main difference is that we could \textit{derive} this result rather than having to postulate it  as  in the case of ideal clocks.
This provides an interesting approach to the principle of equivalence in the quantum domain and is worth pursuing further.
\begin{center}
\begin{table}
\begin{tabular}{ |c|c|c|c|c|c| } 
 \hline
 & ~Inertial~ &  ~ Rindler 1 ~  & ~ Rindler 2 ~& ~ Rindler 3 ~ & ~...~ \\ 
 \hline
$|0\rangle_{M}$ & $T=0$ & $T \sim a_1$ & $T \sim a_2$   & $T \sim a_3$   & .. \\ 
 \hline
 $|0\rangle_{R_1}$ & - & $T=0$  & $T \sim a_2$ & $T \sim a_3$   & .. \\
 \hline
 $|0\rangle_{R_2}$ & - & -  & $T =0 $ &  $T \sim a_3$ &  ..\\ 
 \hline
 $|0\rangle_{R_3}$ & - & -  & -  & $T =0 $ & ..\\ 
 \hline
 ... & - & -  & - &  - & .. \\ 
 \hline
\end{tabular}
\caption{Thermal description of various vacua from the perspectives of observers in different frames, with origins of co-ordinates of the each successive frames are translated by a non-zero amount.} 
\label{Table1}
\end{table}
\end{center}
Moreover,  two different uniformly accelerated observers do not have any thermal correspondence if they asymptote to the same null line, where as two accelerated observers having the same acceleration may also have a (uni directional) thermal relation between them if they asymptote to different null line eventually.

Most interestingly, the thermality between the successive Rindler frames is discontinuous in the shift parameter between the Rindler wedges.  This result indicates that a Rindler frame finds the states of  all its predecessors being thermally populated even for the smallest Planck level shift. This realization may have very interesting implications for horizon shifts by microscopic scales which are unable to be detected by macroscopic or classical measurements. For instance, if a black hole evaporates by emitting a Hawking quanta and its horizon shifts by a microscopic amount, the exterior accelerating  observer who was asymptoting to the erstwhile horizon, may in this setting  find the quantum field thermally populated, non-perturbatively indicating a shift in horizon. Though we have done the explicit computations for a $1+1$ dimensional black hole setting, the causal diamonds for accelerating observers  will still have this nested structure even in the $3+1$ dimensional cases, and one can expect results on the similar lines as the validity of the Eq.(\ref{EffectBeta1}), at least for planar motions in a spherically symmetric spacetime.

These results gives rise to intriguing possibility of a new set of observers in the exterior off the Schwarzschild spacetime which are neither uniformly accelerated nor static observers who still perceive a horizon and receive thermal radiation from it at the semiclassical level. Thus, it opens up a possibility of a new class of observers who may ascribe thermal description to their causal patch. All such issues will be discussed in details elsewhere.

\begin{acknowledgments}
 The authors thank Ulrich Gerlach, Jorma Louko  and Karthik Rajeev for useful comments on a previous version of the manuscript. Research of K.L. is partially supported by the  Government of India through the ANRF grant  MTR/2022/000900. Research of T.P. was partially supported by the J.C.Bose Fellowship of Department of Science and Technology, Government of India.
\end{acknowledgments}


\providecommand{\href}[2]{#2}\begingroup\raggedright\endgroup

\begin{appendix}
\begin{widetext}
\section*{Appendix A : Direct Evaluation of the Bogoliubov coefficients} \label{DirectComp}
For a 1+1 d  spacetime, we write the Rindler spacetime in conformally flat co-ordinate $(\tau,\zeta)$. Let there be two inertial observers $I_1\equiv (t,x)$ and $I_2 \equiv(t',x')$ translated along the spatial axis,
$x' =x-\ell_{1}$. 
 Let there be two Rindler observers  $R_0\equiv(\tau_0,\zeta_0)$ and $R_1\equiv(\tau_1 ,\zeta_1)$ moving with accelerations $a_0$ and $a_1$ respectively with respect to the inertial observers.
In conformally flat co-ordinates, the mode function in the Rindler co-ordinates are
\bea
u_{\omega}^{R_i} = \frac{1}{\sqrt{2 \pi \omega}}e^{-i \omega(\tau_i \pm \zeta_i)} = \left[  \frac{1}{\sqrt{2 \pi \omega}}e^{-i \omega \tilde{u}_i} , \frac{1}{\sqrt{2 \pi \omega}}e^{-i \omega \tilde{v}_i)} \right]
\eea
where $\tilde{u},\tilde{v}$ are the double null co-ordinates in the Rindler frames. 

The relation between the Rindler and inertial co-ordinates are given as follows
\bea
\text{Between $R_0$ and $M_1$ :  } 
t &=& \frac{1}{a_0}e^{a_0 \zeta_0}\sinh{(a_0 \tau_0)}\nonumber\\
x &=& \frac{1}{a_0}e^{a_0 \zeta_0}\cosh{(a_0 \tau_0)};\label{Rind-Iner 1}\\
\text{Between $R_1$ and $M_2$ :  } 
t' &=& \frac{1}{a_1}e^{a_1 \zeta_1}\sinh{(a_1 \tau_1)}\nonumber\\
x' &=& \frac{1}{a_1}e^{a_1 \zeta_1}\cosh{(a_1 \tau_1)}. \label{Rind-Iner 2}
\eea
Using $x' =x-\ell_{1}$  and \eqref{Rind-Iner 2}, we have the two Rindler frames connected as 
\bea
\frac{1}{a_0}e^{a_0 \zeta_0}\sinh{(a_0 \tau_0)}&=& \frac{1}{a_1}e^{a_1 \zeta_1}\sinh{(a_1 \tau_1)}\nonumber\\
\frac{1}{a_0}e^{a_0 \zeta_0}\cosh{(a_0 \tau_0)}-\ell_{1} &=& \frac{1}{a_1}e^{a_1 \zeta_1}\cosh{(a_1 \tau_1)}. \label{Rind_XX}
\eea
Using  \eqref{Rind_XX} we have
\bea
\sinh{(a_1 \tau_1)} = \frac{\sinh{(a_0 \tau_0)}}{\cosh{(a_0 \tau_0)}}\left[ \cosh{(a_1 \tau_1)}+ a_1 \ell_{1} e^{-a_1 \zeta_1} \right].
\eea
Thus,
\bea
\tanh{(a_0 \tau_0)}= \frac{\tanh{(a_1 \tau_1)}}{1 + g \sech{(a_1 \tau_1)}},
\eea
with $g\equiv g(\ell_{1},a_1 \zeta_1)= a_1 \ell_{1} e^{-a_1 \zeta_1} $. Therefore, 
\bea
e^{-i \omega \tau_0} = e^{- \frac{\omega_0}{a_0}\tanh^{-1}\left[\frac{\tanh{(a_1 \tau_1)}}{1 + g \sech{(a_1 \tau_1)}}\right]}.
\eea
Further, from \eqref{Rind_XX}, we have
\bea
e^{a_0 \zeta_0} &=& \frac{a_0}{a_1} e^{a_1 \zeta_1} \frac{\sinh{(a_1 \tau_1)}}{\sinh{(a_0 \tau_0)}},\\
e^{i \omega_0 \zeta_0}                            &=&\left(\frac{a_0}{a_1}\right)^{i \frac{\omega}{a_0}}  e^{i \frac{\omega}{a_0} a_1 \zeta_1} \left( 1 + g \sech{(a_1 \tau_1)}\right)^{i \frac{\omega}{a_0}} \left( \frac{\cosh{(a_1 \tau_1)}}{\cosh{(a_0 \tau_0)}}\right)^{i \frac{\omega}{a_0} }.
\eea 
Using the representation of $\tanh^{-1}{z}=(\log{1+z}-\log{1-z})/2$ for $|z|<1$
\bea
\tanh^{-1}\left[\frac{\tanh{(a_1 \tau_1)}}{1 + g \sech{(a_1 \tau_1)}}\right]=\frac{1}{2} \log{\left[\frac{\cosh{(a_1 \tau_1)} +g +\sinh{(a_1\tau_1)}}{\cosh{(a_1 \tau_1)} +g -\sinh{(a_1\tau_1)}}\right]},
\eea
and
\bea
\frac{\cosh{(a_1 \tau_1)}}{\cosh{(a_0 \tau_0)}}=\frac{[(\cosh{(a_1 \tau_1)} +g)^2 -\sinh^2{(a_1\tau_1)}]^{1/2}}{1 + g \sech{(a_1\tau_1)}}.
\eea
Therefore,
\bea
e^{-i \omega (\tau_0-\zeta_0)} = \left(\frac{a_0}{a_1}\right)^{i \frac{\omega}{a_0}}[\cosh{(a_1 \tau_1)} +g -\sinh{(a_1\tau_1)} ]^{i \frac{\omega}{a_0}} e^{i \frac{\omega}{a_0} a_1 \zeta_1}.
\eea
Using $\tau_1 =(\tilde{u}_1+\tilde{v}_1)/2$ and $\zeta_1 =(\tilde{v}_1-\tilde{u}_1)/2$,
\bea
e^{-i \omega \tilde{u}_0} = \left(\frac{a_0}{a_1}\right)^{i \frac{\omega}{a_0}} e^{-i \frac{\omega}{a_0} a_1 \tilde{u}_1}\left[1 + a_1 \ell_{1} e^{a_1 \tilde{u}_1} \right]^{i \frac{\omega}{a_0}}. \label{R1R2Mode}
\eea
In the region of $R_1$ the mode functions are related as
\bea
u_{\omega}^{R_1} = \sum_{\omega'} [ \tilde{\alpha}_{ \omega,\omega'} u_{\omega'}^{R_0} + \tilde{\beta}_{ \omega,\omega'} (u_{\omega'}^{R_0} )^*].
\eea
Therefore,
\bea
\tilde{\beta}_{\omega, \omega''}&=& - ((u_{\omega''}^{R_0} )^*,u_{\omega}^{R_1} )=((u_{\omega}^{R_1} )^*,u_{\omega''}^{R_0} ) =-\frac{i}{2 \pi \sqrt{\omega \omega''}} \int_{-\infty}^{\infty} d \tilde{u}_1e^{- i \omega'' \tilde{u}_0}\partial_{\tilde{u}_1} e^{- i \omega\tilde{u}_1};\\
&=& - \frac{1}{2 \pi} \sqrt{\frac{\omega}{ \omega''}}  \left(\frac{a_0}{a_1}\right)^{i \frac{\omega''}{a_0}} \int_{-\infty}^{\infty} d \tilde{u}_1  e^{-i \left(\frac{\omega''}{a_0} a_1  + \omega \right)\tilde{u}_1}\left[1 + a_1\ell e^{a_1 \tilde{u}_1} \right]^{i \frac{\omega''}{a_0}};\\
&=&  -\frac{1}{2 \pi a_1} \sqrt{\frac{\omega}{ \omega''}}  \left(\frac{a_0}{a_1}\right)^{i \frac{\omega''}{a_0}}(a_1 \ell_{1})^{i\left( \frac{\omega''}{a_0} + \frac{\omega}{a_1}\right)}\frac{\Gamma\left[i \frac{\omega}{a_1} \right] \Gamma\left[-i\left( \frac{\omega''}{a_0} + \frac{\omega}{a_1}\right) \right]}{\Gamma\left[-i \frac{\omega''}{a_0} \right]};\\
&=& - \frac{1}{2 \pi^2} \frac{\sqrt{\omega \omega''}}{a_0 a_1}  \left(\frac{a_0}{a_1}\right)^{i \frac{\omega''}{a_0}}(a_1 \ell)^{i\left( \frac{\omega''}{a_0} + \frac{\omega}{a_1}\right)}\Gamma\left[\frac{i\omega''}{a_0} \right]\Gamma\left[\frac{i\omega}{a_1} \right] \Gamma\left[-i\left( \frac{\omega''}{a_0} + \frac{\omega}{a_1}\right) \right]\sinh{\left(\frac{\pi \omega''}{a_0}\right)}, \label{EffBogBeta2}
\eea
where the inner product $((u_{\omega}^{R_1} )^*,u_{\omega''}^{R_0} )$  in the first step gets effectively evaluated inside the region  $R_1$ as $u_{\omega}^{R_1}$ vanishes outside. The above expression exactly matches Eq.(11) of the main paper, leading to
\bea \label{NumExp}
N_{\omega} =\frac{1}{4 \pi a_1} \frac{1}{\sinh{\frac{\pi \omega}{a_1}}}\int_0^{\infty} \frac{d \omega''}{a_0} \frac{\sinh{\frac{\pi \omega''}{a_0}}}{\left( \frac{\omega}{a_1}+\frac{\omega''}{a_0} \right)\sinh{\left( \frac{\omega}{a_1}+\frac{\omega''}{a_0} \right)}}.
\eea
For no translation , i.e. $\ell_{1} =0$, we see from  Eq. \eqref{Rind_XX} 
\bea
\tanh{a_0 \tau_0} &=& \tanh {a_1 \tau_1} \Rightarrow a_0\tau_0 =a_1 \tau_1 +  i n\pi, \hspace{0.2 in} \text{and, }\\
\frac{e^{a_0 \zeta_0}}{e^{a_1 \zeta_1}} &=&\pm \frac{a_0}{a_1}.
\eea
Therefore,
\bea
e^{-i \omega \tilde{u}_0} = e^{-i \omega(\tau_0 -\zeta_0)} = e^{-i \omega \frac{a_1}{a_0}\tilde{u}_1} \left( \pm \frac{a_0}{a_1}\right)^{\frac{i \omega }{a_0}}e^{- n \pi  \frac{\omega }{a_0}} = e^{-i \omega \frac{a_1}{a_0}\tilde{u}_1} \left(\frac{a_0}{a_1}\right)^{\frac{i \omega }{a_0}}e^{- 2 n \pi  \frac{\omega }{a_0}},
\eea
also obvious from Eq. \eqref{R1R2Mode}.
Hence the Bogoliubov coefficient
\bea
\tilde{\beta}_{\omega, \omega''}&=& -\frac{i}{2 \pi \sqrt{\omega \omega''}} \int_{-\infty}^{\infty} d \tilde{u}_1e^{- i \omega'' \tilde{u}_0}\partial_{\tilde{u}_1} e^{- i \omega\tilde{u}_1} \sim \int_{-\infty}^{\infty} d \tilde{u}_1  e^{- i \left(\omega +  \frac{a_1}{a_0} \omega''\right)\tilde{u}_1},\\
&\sim& \delta \left( \omega +  \frac{a_1}{a_0} \omega'' \right) =0,
\eea
which reinforces Eq.(3) of the main paper.
\section*{Appendix B: Expectation value of the Number Operator}\label{NumberCount}
Taking $\omega''/a_0 =x$ and $\omega/a_1=y$, we break the integral in Eq. \eqref{NumExp} into two parts as \\
\bea
{\bf I} =\int_0^{\infty} dx \frac{e^{\pi x}}{(x+y) \sinh{(x+y)}} &=& \int_y^{\infty}dz\frac{e^{\pi(z-y)}}{z\sinh{z}}=2e^{-\pi y} \int_y^{\infty} \frac{dz}{z}\frac{1}{1-e^{-2 \pi z}};\\
&=& 2e^{-\pi y} \sum_{n=0}^{\infty}\int_y^{\infty} \frac{dz}{z}e^{-2 n \pi z} =  2e^{-\pi y}\int_y^{\infty} \frac{dz}{z} +  2e^{-\pi y} \sum_{n=1}^{\infty}\int_y^{\infty} \frac{dz}{z}e^{-2 n \pi z};\nonumber\\
&=&  2e^{-\pi y} \int_y^{\infty} \frac{dz}{z} +  2e^{-\pi y}\sum_{n=1}^{\infty}\Gamma\left(0,2 n \pi y \right).
\eea
 Similarly,
\bea
{\bf II} = \int_0^{\infty} dx \frac{e^{-\pi x}}{(x+y) \sinh{(x+y)}} &=& \int_y^{\infty}dz\frac{e^{-\pi(z-y)}}{z\sinh{z}}=2e^{\pi y} \int_y^{\infty} \frac{dz}{z}\frac{e^{-2 \pi z}}{1-e^{-2 \pi z}};\\
&=&   2e^{\pi y} \sum_{n=1}^{\infty}\int_y^{\infty} \frac{dz}{z}e^{-2 n \pi z} = 2e^{\pi y}\sum_{n=1}^{\infty}\Gamma\left(0, 2 n \pi y\right).
\eea
Therefore,
\bea
\int_0^{\infty} dx \frac{\sinh{\pi x}}{(x+y) \sinh{(x+y)}} = e^{-\pi y} \int_y^{\infty} \frac{dz}{z} -  2\sinh{\pi y}\sum_{n=1}^{\infty}\Gamma\left(0,2 n \pi y\right). \label{NumberIntegral}
\eea
\subsubsection*{Separating the divergent part in the  Integral}
\noindent The first integral on the RHS of Eq. \eqref{NumberIntegral} is of the form 
\bea
\int_{\frac{\omega}{a_2}}^{\infty}\frac{d z}{z} = \int_{\log{\frac{\omega}{a_2}}}^{\infty}dt.
\eea
Now,
\bea
\int_b^a dt e^{i \omega t} =-i\left[ -\frac{2 \sin^2{\frac{\omega a}{2}}}{\omega}  +\frac{2 \sin^2{\frac{\omega b}{2}}}{\omega}\right] + \frac{\sin{\omega a}}{\omega}-\frac{\sin{\omega b}}{\omega}.
\eea
For $a \rightarrow \infty$, we have
\bea
\int_b^{\infty} dt e^{i \omega t} =-i\left[ -\frac{\pi^2}{2} \omega (\delta(\omega))^2  +\frac{2 \sin^2{\frac{\omega b}{2}}}{\omega}\right] + \pi \delta(\omega)-\frac{\sin{\omega b}}{\omega}.
\eea
Now for $\omega \rightarrow 0$ we have
\bea
\int_b^{\infty} dt  =-i\left[ -\frac{\pi^2}{2}\lim_{\omega \rightarrow 0} \omega (\delta(\omega))^2  \right] + \pi \delta(0)- b.
\eea
The first term in the RHS vanishes for any regular distribution. Also it vanishes if we take $\omega \rightarrow 0$ limit first, thus in combination with Eqs. \eqref{NumExp} and \eqref{NumberIntegral} obtaining 
\bea
N_{\omega}= \frac{1}{2 \pi a_1}\left[ \frac{\pi \delta(0)}{e^{\frac{2 \pi \omega}{a_1}}-1}  - \frac{\log{\frac{\omega}{a_1}}}{e^{\frac{2 \pi \omega}{a_1}}-1}-\underbrace{ \sum_{n=1}^{\infty}\Gamma \left( 0, 2  n \frac{\pi \omega}{a_1} \right)}_{\text{Convergent series}}\right] \approx  \frac{1}{(2 \pi a_1)}\frac{\pi \delta(0)}{e^{\frac{2 \pi \omega}{a_1}}-1} .
\eea
We can see that $\delta(0)$ identifies the volume of the ``box'' where this computation is carried out. Thus, in a free space it infinitely overdominates the finite subleading correction terms, effectively leading to Eq. (5) of the main paper.  Further, the higher correlations of the number operator can be obtained from here and following the arguments of \cite{Fabbri:2005} one can show that due to similar suppressions of the correction terms, the higher correlations also maintain a thermal character, i.e.
\bea
\langle 0_{R_0}|\hat{N}_{\omega_1}\hat{N}_{\omega_2}|0_{R_0}\rangle &=& N_{\omega_1}N_{\omega_2} + \left|\int_0^{\infty} d \omega' \tilde{\beta}_{\omega_1, \omega'} \tilde{\beta}^*_{\omega_2, \omega'}\right|^2 +  \left|\int_0^{\infty} d \omega' \tilde{\alpha}_{\omega_1, \omega'} \tilde{\beta}_{\omega_2, \omega'}\right|^2 \label{Correlator}\\
&\approx&  N_{\omega_1}N_{\omega_2} \approx  \frac{1}{(2 \pi a_1)^2}\frac{\pi \delta(0)}{e^{\frac{2 \pi \omega_1}{a_1}}-1} \frac{\pi \delta(0)}{e^{\frac{2 \pi \omega_2}{a_1}}-1},
\eea
with the last two terms  in Eq.(\ref{Correlator}) subdued by the dominant $ N_{\omega_1}N_{\omega_2}$ term. Further, looking at the structure of  $\tilde{\beta}_{\omega,  \omega'}$ and $\tilde{\alpha}_{\omega, \omega'}= - i\tilde{\beta}_{\omega,- \omega''}$ it is easy to verify that like the number spectrum such correlators are also free from $\ell $ or $a_0$.
So, as far as the distribution profile is concerned, the quantum field in the vacuum of $R_0$ appears to be thermally populated in the
 frame $R_1$ with absolutely no reference to $\ell$ or the label $a_0$ of the frame whose vacuum is being considered here. Interestingly, in addition, there may exist other class of correlators which may contain the information of the parameter $\ell_1$ \cite{Lochan:2016nbs}.


\section*{Appendix C : Periodicity in the correlation functions}\label{Cyclicity}
In the natural vacua of $R_0$, i.e. $|0\rangle_{R_0}$ the Wightman function of the scalar field can be expressed in terms of the null co-ordinates of $R_0$ as
\bea
G_{\phi}=\int d\omega u_{\omega}^{R_0}(\tau_0,\zeta_0) u_{\omega}^{R_0*}(\tau_0',\zeta_0') = \int \frac{d\omega}{2 \omega}e^{-i \omega \tilde{u}_0}e^{i \omega \tilde{u}_0'}.
\eea
Using Eq.\eqref{R1R2Mode} this correlation function as seen in the frame $R_1$ appears as
\bea
G_{\phi}\sim\int \frac{d\omega}{2 \omega} \left( e^{a_1( \tilde{u}_1-\tilde{u}_1')} \right)^{ -i \frac{\omega}{a_0}}\left[\frac{1 + a_1 \ell_{1} e^{a_1 \tilde{u}_1}}{1 + a_1 \ell_{1} e^{a_1 \tilde{u}_1'}}  \right] ^{i \frac{\omega}{a_0}}.
\eea
Clearly we can see there exists a periodicity in $G_{\phi}$ with respect to $\tau_1\rightarrow \tau_1 +2 i n \pi/a_1$ and 
$\tau_1' \rightarrow \tau_1' +2 i  m \pi/a_1$ for integer $n,m$ since this corresponds to  $\tilde{u}_1\rightarrow \tilde{u}_1 +2 i  n \pi/a_1$ and $\tilde{u}_1'\rightarrow \tilde{u}_1' +2 i  m \pi/a_1$. This periodicity is on the same footing a generic quantum state of the inertial Fock space offers to the correlation functions when depicted in the frame $R_0$, for instance. Therefore, an equivalent thermal correspondence exists between $R_0$ and $R_1$ as well.
\end{widetext}
\end{appendix}
\end{document}